\documentclass[preprint,showpacs,preprintnumbers,amsmath,amssymb,floats,nofootinbib]{revtex4}
\usepackage[dvips]{graphicx}
\usepackage[english]{babel}
\usepackage{amsmath}
\usepackage{amssymb}
\usepackage{bm}

\newcommand{\vk}{\mathbf{k}}

\newcommand{\vR}{\mathbf{R}}
\newcommand{\us}{\uparrow}
\newcommand{\ds}{\downarrow}
\newcommand{\be}{\begin{eqnarray}}
\newcommand{\ee}{\end{eqnarray}}
\newcommand{\p}{\partial}
\newcommand{\cd}{c^{\dagger}}

\def\ket#1{|#1\rangle}
\def\bra#1{\langle #1 |}
\def\ep#1{\langle #1 \rangle}

\begin{document}

\title{Quantum Spin Hall Effect in a Three-orbital Tight-binding Hamiltonian}
\author{Yan He$^1$ and Changtao Hou$^2$}

\affiliation{$^1$Department of Physics, University of Chicago, Chicago, IL\\
$^2$Department of Physics, University of California, Riverside, CA}
\date{\today}

\begin{abstract}
We consider the quantum spin hall state in a three orbital model due to certain loop current order induced by spin-dependent interactions. This type of order is motivated by the loop current model which is proposed to describe the pseudogap phase of cuprates. It is shown that this model has nontrivial Chern parity by directly counting the zeros of the Pfaffian of time reversal operator. By connecting to the second order Chern number, we explicitly show the singularities of wave-functions and how they depend on gauge choices. In this case, it is shown that the Berry phase can be mapped to Nonabelian instanton.
\end{abstract}

\maketitle

\section{Introduction}

In recent years, there are a lot of interests arise in the topological property of band structures, such as quantum anomalous Hall effects (QAH) and quantum spin Hall (QSH). All these topological nontrivial states are inspired by the integer quantum hall effects, in which it is shown that the quantized hall conductance is also the Chern number of underlying band structure \cite{tknn}. Haldane introduced a complex next nearest neighbor hopping in a 2D lattice model to give rise a nonzero Chern number without applying an external magnetic field \cite{haldane}. This leads to an extensive study on QAH \cite{haldaneahe,nagaosaahereview}. It was also shown that Chern number is the only topological invariant which characterize time reversal breaking Hamiltonian \cite{avronseilersimon}.

In our previous paper \cite{yanheAHE}, it is noted that Haldane's model can be realized in the lattice model with some special time reversal breaking loop current ordered states due to the nearest neighbor interaction. The loop current ordered states without changing translational symmetry were originally proposed to arise as broken symmetry states due to interactions in a three-orbital model for underdoped cuprates \cite{cmv97, simon-cmv, cmv06} and have been discovered in several families of cuprates \cite{bourges, kaminski, li}. These observed loop-current states, however, do not lead to the quantized anomalous Hall effect (QAHE) or nonzero Chern number because they preserve the product of time reversal and spatial inversion symmetry. But other possible type of loop current order will lead to QAH.  Also in \cite{yanheAHE}, the discussion of the singularities of wave function helps to reveal the topological obstruction in QAH states.

In this paper, we generalize these results to include spin degree of freedom and consider the ordered spin loop current states due to the nearest neighbor interaction. This will restore the time reversal symmetry and also make the Chern number trivially zero. Although the first order Chern number is zero, it is pointed out that the time reversal invariant Hamiltonian is characterized by the second order Chern number \cite{Simon}. Many yeas later, in the seminal paper \cite{kane-mele}, Kane and Mele introduced the $Z_2$ invariant to classify the two dimensional time reversal invariant model, which also lead to the suggestion of 2D and 3D topological insulator \cite{hasankane,moorenature,qizhangreview}. These two classification of time reversal invariant model seem to be contradicted to each other. This contradiction can be resolved by the dimension reduction method as in \cite{zhang}. Another more geometric method to connect the Chern parity to the second Chern number can be found in \cite{hatsugai}. This argument is very similar to the one used in the Witten's $SU(2)$ global anomaly \cite{Witten}. We will first follow Kane and Mele to construct the $Z_2$ invariant directly. After connecting the $Z_2$ invariant to the second Chern number, we will analyze the singular points of the wave function of a two band model example in detail just as in the QAH case \cite{yanheAHE}. For the two band time reversal breaking Hamiltonian, it is well know that the $U(1)$ Berry phase configuration looks like a monopole solution \cite{nakahara}. Similarly, for the two band time reversal invariant Hamiltonian, we found the $SU(2)$ Berry phase configuration can be mapped to the self-adjoint instanton solution of Non-abelian gauge theory after some suitable coordinate transformations. At last, the three orbital Cu-O lattice model will be discussed. Again we will directly construct the $Z_2$ invariant to verify it is a  QSH state.

\section{spin loop current states in the Copper-oxygen lattice model}
\label{spin}

\begin{figure}
\centerline{\includegraphics[width=0.4\textwidth]{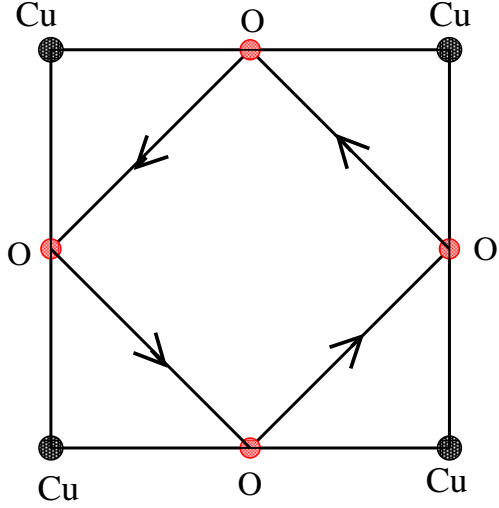}}
\caption{Schematic figure of loop current in Cu-O lattice}
\label{fig:cuo}
\end{figure}

We consider the two-dimensional lattice with the structure of the copper oxides lattice as shown in Fig. (\ref{fig:cuo}). There are three orbitals per unit-cell, which are the $d$-orbital on the copper atom and the $p_x$ and $p_y$ orbitals on the oxygens.  The minimal kinetic energy operator with a choice of gauge such that the $d$ orbital is purely real and the $p_x$ and the $p_y$ orbitals purely imaginary is
\be
\label{ke}
H_{KE}=it\,d^{\dagger}_{\vk}(s_xp_{x,\vk}+s_yp_{y,\vk})
-t's_xs_yp_{x,\vk}^{\dagger}p_{y,\vk}+h.c.
\ee
with $s_x=\sin(k_x/2)$ and $s_y=\sin(k_y/2)$ for a lattice constant taken to be 1.

In the loop current model of pseudogap phase, one usually only consider the charge channel, thus is equivalent to consider spinless fermions. To consider the spin Hall effects, we also need to decompose the interaction term in the spin channel. As we know from the previous paper, the only topological nontrivial state is obtained by considering the interaction between the $p$ orbitals as
\be
\label{int}
H_{int}=\sum_{\ep{i,j}}\sum_{\sigma,\sigma'}Vn_{p,i,\sigma}n_{p,j,\sigma'}.
\ee
Here $i,j$ labels the lattice sites and $\sigma,\sigma'$ labels the spin.

In the spinless case, the the above interaction term can be decomposed by using the operator identity $n_i n_j = -\frac{1}{2}(|J_{ij}|^2-n_i-n_j)$ with charge current operator $J_{ij} = i(c_i^{\dagger}c_j - c_j^{\dagger}c_i)$. In the mean field theory, one of current in the current interaction term can be replace by its expectation as $(V/2)\langle J_{ij}\rangle =i r$, thus one finds an interaction induced kinetic energy term
\be
H_{int}'=irc_xc_yp_{x,\vk}^{\dagger}p_{y,\vk}+h.c.
\ee
If $r\neq0$ is a stable state, it describes loop currents flowing clockwise (or anti-clockwise) around the oxygen's in each unit-cell as shown in Figure (\ref{fig:cuo}). This is one of the five possible loop-curent states with non-overlapping loops in the Cu-O lattice all of which preserve translational symmetry. In Figure (\ref{fig:cuo}), the flux has one sign in the square formed by the nearest neighbor oxygens which surround a Cu and the opposite sign in the square formed by the nearest neighbor oxygens which do not surround a Cu. Therefore the total flux is zero.

For the spinful fermions, the above operator identity can be trivially generalized to the interaction term which are diagonal in spin indices. Following similar method, the density coupling can be rewritten as spin currents coupling by operator identity $n_{i,\sigma} n_{j,\sigma}
=-\frac12\big(|j_{ij,\sigma}|^2-n_{i,\sigma}-n_{j,\sigma}\big)$ where the spin current is $J_{ij,\sigma} = i(c_{i,\sigma}^{\dagger}c_{j,\sigma} - c_{j,\sigma}^{\dagger}c_{i,\sigma})$.

For the off-diagonal terms in spin indices such as $n_{i,\us}n_{j,\ds}+n_{j,\us}n_{i,\ds}$, the decomposition is more complicated. We can begin with the product of the spin current of two different spins.
\be
J_{ij,\us}J_{ji,\ds}&=&-\cd_{i\us}c_{j\us}\cd_{j\ds}c_{i\ds}-\cd_{j\us}c_{i\us}\cd_{i\ds}c_{j\ds}
+\cd_{i\us}c_{j\us}\cd_{i\ds}c_{j\ds}+\cd_{j\us}c_{i\us}\cd_{j\ds}c_{i\ds}\nonumber\\
&=&S^+_iS^-_j+S^-_iS^+_j-T^+_iT^-_j-T^-_iT^+_j
\ee
Here $S^+_i=\cd_{i\us}c_{i\ds}$, $S^-_i=\cd_{i\ds}c_{i\us}$ and $S^z_i=\frac12(n_{i,\us}-n_{i,\ds})$ form the spin $SU(2)$ algebra, and $T^+_i=\cd_{i\us}\cd_{i\ds}$, $T^-_i=c_{i\ds}c_{i\us}$ and $T^z_i=\frac12(n_{i,\us}+n_{i,\ds}-1)$ form the charge $SU(2)$ algebra. Making use of the following identities
\be
&&{\bf S}_i\cdot{\bf S}_j=\frac12(S^+_iS^-_j+S^-_iS^+_j)+S^z_iS^z_j\\
&&{\bf T}_i\cdot{\bf T}_j=\frac12(T^+_iT^-_j+T^-_iT^+_j)+T^z_iT^z_j
\ee
we find the following decomposition.
\be
n_{i,\us}n_{j,\ds}+n_{j,\us}n_{i,\ds}=J_{ij,\us}J_{ji,\ds}
-2({\bf S}_i\cdot{\bf S}_j-{\bf T}_i\cdot{\bf T}_j)+\frac12(n_i+n_j)
\ee
Other than the spin current couplings, there are also more complicated terms like spin and charge operator interaction between neighboring sites. Since we only want to consider current order here, the spin and charge interaction term will be ignored. As before, in the mean field level, the current can be replaced by its expectations, then we find a new kinetic term
\be
H_{int}'=ir_{\sigma}c_xc_yp_{x,\vk,\sigma}^{\dagger}p_{y,\vk,\sigma}+h.c.
\ee
If the expectation of spin up and spin down current are opposite, this loop current ordered system is time reversal invariant again and is possible to give rise a quantum spin hall states.

We will consider the QSH state of the 3 orbital Cu-O model and therefore the singularities of wave functions of the model with the Hamiltonian $H=H_{KE}+H_{int}'$. Before we do that, let us first consider the simpler case of a two band model.

\section{Two band model with spin}

In \cite{yanheAHE}, we have discussed a simple two-band Hamiltonian on a square lattice from a wave function point of view.
\be
H=\vR\cdot\bm{\sigma},\quad\vR=(m+\cos k_x+\cos k_y,\,\sin k_x,\,\sin k_y)
\label{m1}
\ee
This Hamiltonian breaks time reversal. For $0<m<2$ or $-2<m<0$, the Chern number is $+1$ or $-1$ respectively for the lower band. In the two band model, the Chern number simply equals to the winding number of the mapping $\vR/|\vR|$ from a 2D torus $T^2$ to a 2D sphere $S^2$. In this case, geometric meaning of Chern number is quite clear. If we change one of the $R_i$ to be $-R_i$, then the orientation of the above mapping is reversed, then Chern number will change sign.

Now we can introduce spin degree of freedom to enlarge the above Hamiltonian to become a 4-band model and restore time reversal symmetry. Then we can show this model has non trivial Chern parity or  Quantum Spin Hall (QSH) effects. The only time reversal breaking term in Eq (\ref{m1}) is $R_3$. The time reversal is restored by multiply a spin operator, thus we find the following model
\be
H=\vR\cdot\bm{\sigma},\quad\vR=(m+\cos k_x+\cos k_y,\,\sin k_x,\,a+s_z\sin k_y)
\ee
Here $s_z$ is the z component of spin. For $s_z=1$ or spin up, we have Chern number equals to 1 and for
$s_z=-1$ or spin down, we have Chern number equals to -1. Here we also insert an arbitrary constant $a$ there in order to lift the degeneracy of the spin up and down bands. In matrix form, the model is
\be
H=R_1(\sigma_1\otimes I)+R_2(\sigma_2\otimes I)+a(\sigma_3\otimes I)+\sin k_x(\sigma_3\otimes s_z)
\ee
Here $\sigma_{1,2,3}$ and $s_{x,y,z}$ are Pauli matrices and $I$ is 2 by 2 identity matrix, $R_1=m+\cos k_x+\cos k_y$ and $R_2=\sin k_y$.

The spin up and spin down two bands are still degenerate at $k_y=0,\pm\pi$. So one cannot distinguish the Chern number between the spin up band and spin down band. The total Chern number is zero. But for certain choice of parameters, there is also $Z_2$ topological invariant. There are several different way to get this $Z_2$ invariant. Here we use the approach proposed in Kane and Mele's original proposal \cite{kane-mele}.
In this method, we define the pfaffian $P(\vk)=\mbox{Pf}[\bra{u_i(\vk)}\Theta\ket{u_j(\vk)}]$. The zeros of $P(\vk)$ always come in pairs, if $\vk$ is a zero point, so do $-\vk$. We can count the number of the pairs. Since two pairs can annihilate each other, even number of pairs are topological trivial, odd number of pairs is QSH state.

In our model, we define the pfaffian $P(\vk)=\bra{u_{\ds}(\vk)}\Theta\ket{u_{\us}(\vk)}$. Here $u_{\us,\ds}$ are the eigenstates of spin up and spin down for the lower two bands. $\Theta=I\otimes(is_y)K$ is the time reversal operator and $K$ is the operator to take complex conjugate. For concreteness, we take $m=-1$ and $|a|<1$ and define $R_{3,\us}=a+\sin k_y$ and $R_{3,\ds}=a-\sin k_y$ and $R_{\us,\ds}=(R_1^2+R_2^2+R_{3,\us,\ds}^2)^{1/2}$. Since $\Theta$ will always change spin up to spin down, we can omit the spin part of wave function and only write out the orbital part of wave function explicitly.

We already know the wave functions, one way to write them is
\be
\ket{u_{\us}}=\frac{1}{\sqrt{2R_{\us}(R_{\us}-R_{3,\us})}}\left(\begin{array}{c}
R_{3,\us}-R_{\us}\\
R_1+iR_2
\end{array}\right)\qquad
\ket{u_{\ds}}=\frac{1}{\sqrt{2R_{\ds}(R_{\ds}+R_{3,\ds})}}\left(\begin{array}{c}
R_1-iR_2\\
-R_{3,\ds}-R_{\ds}
\end{array}\right)
\label{wf}
\ee
The above wave functions are well defined if $R_1$ and $R_2$ are not equal to zero at the same time.

First we consider a special case $R_1=R_2=0$. We will that show that if $R_1=R_2=0$, then $P(\vk)=0$. $R_1=R_2=0$ only happens for $k_x=0$ and $k_y=\pm\frac{\pi}{2}$. At these two points, Eq.(\ref{wf}) may be ill defined. In this case, it is easy to find the eigenstates if we go back to the Hamiltonian. The Hamiltonian for spin up and spin down at these points are actually diagonal matrices
\be
H_{\us}=\left(\begin{array}{cc}
(a+\sin k_y) & 0\\
0 & -(a+\sin k_y)
\end{array}\right)\qquad
H_{\ds}=\left(\begin{array}{cc}
(a-\sin k_y) & 0\\
0 & -(a-\sin k_y)
\end{array}\right)
\ee
At these two point $\sin k_y=\pm1$. For example, for $k_x=0$, $k_y=\pi/2$, then we find
$\ket{u_{\us}}={0 \choose 1}$ and $\ket{u_{\ds}}={1 \choose 0}$, from this clearly $P(\vk)=0$. Similar thing also happens at $k_x=0$, $k_y=-\pi/2$.

So far we have found one pair of zeros. Now we want to show there is no other zeros. Thus this is the only pair of zeros and the number is odd, so we have QSH state. Since we are only looking for zeros, we can drop the normalization factors. For cases other than $R_1=R_2=0$, Eq. (\ref{wf}) is well defined, thus we find the pfaffian as
\be
\bra{u_{\ds}}\Theta\ket{u_{\us}}&\propto&
(R_1+iR_2,\,-R_{3,\ds}-R_{\ds})\cdot\left(\begin{array}{c}
R_{3,\us}-R_{\us}\\
R_1-iR_2
\end{array}\right)\nonumber\\
&=&R_1(R_{3\us}-R_{3\ds}-R_{\us}-R_{\ds})+iR_2(R_{3\us}+R_{3\ds}-R_{\us}+R_{\ds})
\ee

Now we show if we assume $\bra{u_{\ds}}\Theta\ket{u_{\us}}=0$, then it will lead to some contradictions. We can distinguish 3 different cases. Note that we always have the following identity
\be
R_{\us}^2-R_{\ds}^2=R_{3,\us}^2-R_{3,\ds}^2
\label{r1}
\ee

First, if $R_1=0$ and $R_2\ne0$, then $\bra{u_{\ds}}\Theta\ket{u_{\us}}=0$
implies $R_{\us}-R_{\ds}=R_{3,\us}+R_{3,\ds}$. Combining with Eq. (\ref{r1}), we find
$R_{\us}=R_{3,\us}$ and $R_{\ds}=-R_{3,\ds}$. But since $R_2\ne0$ ,we should have $|R_{\us}|>|R_{3\us}|$ and $|R_{\ds}|>|R_{3\ds}|$ which is contradict with above equations.

Similarly, If $R_1\ne0$ and $R_2=0$, then $\bra{u_{\ds}}\Theta\ket{u_{\us}}=0$ implies $R_{\us}+R_{\ds}=R_{3,\us}-R_{3,\ds}$. Combining with Eq. (\ref{r1}), we find
$R_{\us}=R_{3,\us}$ and $R_{\ds}=-R_{3,\ds}$ Again since $R_1\neq0$, we should have $|R_{\us}|>|R_{3\us}|$ and $|R_{\ds}|>|R_{3\ds}|$ which is contradict with above equations.

At last, If both $R_1\neq0$ and $R_2\neq0$, then $\bra{u_{\ds}}\Theta\ket{u_{\us}}=0$ implies
$R_{\us}+R_{\ds}=R_{3,\us}-R_{3,\ds}$ and $R_{\us}-R_{\ds}=R_{3,\us}+R_{3,\ds}$,
which means that $R_{\us}=R_{3,\us}$ and $R_{\ds}=-R_{3,\ds}$. Again since $R_1\ne0$ and $R_2\ne0$, we should have $|R_{\us}|>|R_{3\us}|$ and $|R_{\ds}|>|R_{3\ds}|$ which is contradict with above equations. Therefore, in all the above 3 cases other than $R_1=R_2=0$, we always find $P(\vk)\neq0$. Thus $k_y=0$, $k_x=\pm\pi/2$ is the only pair of zeros.

The $Z_2$ invariant of this two band toy model can be determined by a much easier formula proposed by Fu and Kane, provided the constant $a$ is zero. In this case, the model is also invariant under the spatial inversion, then the $Z_2$ invariant is the product of the eigenvalues of the spatial inversion operator at the 4 time reversal invariant momentum (TRIM) points. We assume that the pauli matrices in our model describe pseudo-spins, and the spatial inversion will exchange the two sublattices. Under these assumption, the inversion is given by the operator $P=\sigma_x\otimes I$. It is easy to verify that $H=PHP$. The eigenvalues of $P$ is sgn$(R_1)$ at the 4 TRIM points, which are one $+1$ and three $-1$ or vice versa. Therefore, the $Z_2$ invariant is $-1$ which implies a topological nontrivial QSH states.

\section{Relation to second Chern number and singular points of wave function}

The above $Z_2$ invariant is directly constructed from the matrix element of time reversal operator $\Theta$. But its geometric meaning is not very clear comparing with the Chern number of the time reversal breaking Hamiltonian. As pointed out by B. Simon many years ago, the topological invariant of  time reversal invariant Hamiltonian system is the second order Chern number \cite{Simon}. The nonzero second order Chern number corresponds to the nontrivial elements of $\pi_3(sp(2))=\pi_3(S^3)=Z$. This result seems contradict with the $Z_2$ classification. Actually, the second order Chern number only works for a Hamiltonian in 4D. If applying it to lower dimensions, one can treat extra spatial coordinates as adiabatic changing parameters. Following the similar arguments as Witten's $SU(2)$ global anomaly \cite{hatsugai}, one can show that for lower dimensional models the 2nd order Chern number reduces to a $Z_2$ invariant and the topological nontrivial one corresponds to odd 2nd order Chern number, which also corresponds to the nontrivial elements of $\pi_4(sp(2))=\pi_4(S^3)=Z_2$.

In the more general 4D model, the geometric meaning of the topological invariant can be understood much easier.  In this case, the nonzero Chern number is also the topological obstruction which prevent us to define the wave function and Berry phase globally on the base manifold. We can understand this in a exact parallel way as in the first order Chern number of QAH state. To this ends, we consider the following model.
\be
H=\sum_{i=1}^5R_i\Gamma_i
\ee
with $\Gamma_i=\sigma_1\otimes s_i$ for $i=1,2,3$ and $\Gamma_4=\sigma_2\otimes I$,  $\Gamma_5=\sigma_3\otimes I$. The connection with the model of Eq (\ref{m1}) discussed above can be obtained by taking
\be
\vR=\Big(\sin k_x,\,\sin k_1, \sin k_2, \sin k_y, m+\cos k_x+\cos k_1+\cos k_2+\cos k_y\Big)
\ee
and then treat the the momentum $k_{1,2}$ as adiabatic parameters.

The energy bands are $E=\pm R$ with double degeneracy. Here $R=\sqrt{\sum_{i=1}^5R_i^2}$. To be explicit, we only consider the lower two degenerate bands. The wave function are given by
\be
&&\ket{\psi_1}=\frac{1}{\sqrt{2R(R+R_5)}}\Big(-R_3+iR_4,\,-R_1-iR_2,\,R_5+R,\,\,0\Big)^T\\
&&\ket{\psi_2}=\frac{1}{\sqrt{2R(R+R_5)}}\Big(-R_1+iR_2,\,R_3+iR_4,\,0,\,\,R_5+R\Big)^T
\ee
For convenience, we perform the calculations in terms of differential forms.
Then the non-abelian Berry phase is $A_{ij}=\bra{\psi_i}d\ket{\psi_j}$, which is a $su(2)$ Lie algebra valued one form. $A_{ij}$ is traceless, thus can be expanded by Pauli matrices as $A=\sum_{a=1}^3A^a\sigma_a/2$. Then we find
\be
A^1=i\frac{R_4dR_1-R_1dR_4+R_2dR_3-R_3dR_2}{R(R+R_5)}\\
A^2=i\frac{R_3dR_1-R_1dR_3+R_4dR_2-R_2dR_4}{R(R+R_5)}\\
A^3=i\frac{R_1dR_2-R_2dR_1+R_4dR_3-R_3dR_4}{R(R+R_5)}
\ee
The Berry curvature is defined as $F=dA+A\wedge A$. Here $F$ is also traceless, thus Tr$F=0$ and the first order Chern number $\int\mbox{Tr}F=0$ as required by the time reversal invariance. Expanding $F$ by Pauli matrices as $F=\sum_aF^a\sigma_a/2$, we find the following
\be
&&F^1=-\frac{R+R_5}{R}A^2\wedge A^3+\frac{R_5}{R^3}(dR_4\wedge dR_1+dR_2\wedge dR_3)
-i\frac{(R+R_5)A^1\wedge dR_5}{R^2}\\
&&F^2=-\frac{R+R_5}{R}A^3\wedge A^1+\frac{R_5}{R^3}(dR_3\wedge dR_1+dR_4\wedge dR_2)
-i\frac{(R+R_5)A^2\wedge dR_5}{R^2}\\
&&F^3=-\frac{R+R_5}{R}A^1\wedge A^2+\frac{R_5}{R^3}(dR_1\wedge dR_2+dR_4\wedge dR_3)
-i\frac{(R+R_5)A^3\wedge dR_5}{R^2}
\ee
The the second order Chern number is related to the winding number of mapping from $T^4$ to $S^4$.
\be
c_2&=&\frac{1}{8\pi^2}\int\mbox{Tr}(F^2)=\frac{1}{8\pi^2}\int\mbox{Tr}\sum_a(F^aF^a)\nonumber\\
&=&-\frac{3}{8\pi^2}\int\frac{1}{R^5}\sum_{i=1}^5(-1)^{i-1}R_idR_1\cdots
dR_{i-1}dR_{i+1}\cdots dR_{5}=-1
\ee
To ease the notations, we omit the $\wedge$ from now on.

One can see that $\ket{\psi_{1,2}}$ is ill defined at the points satisfying $R_i=0$ for $i=1,\cdots,4$ and $R_5<0$. To cover the whole manifold we can introduce another wave function with different gauge choice.
\be
&&\ket{\psi_1}^g=\frac{1}{\sqrt{2R(R-R_5)}}\Big(R-R_5,\,0,\,-R_3-iR_4,\,-R_1-iR_2\Big)\\
&&\ket{\psi_2}^g=\frac{1}{\sqrt{2R(R-R_5)}}\Big(0,\,R-R_5,\,-R_1+iR_2,\,R_3-iR_4\Big)
\ee
$\ket{\psi_{1,2}^g}$ are ill defined at the points satisfying $R_i=0$ for $i=1,\cdots,4$ and $R_5>0$. Other than these points, both of them are well defined and are related to each other by a gauge transformation.
\be
\ket{\psi_i}^g=g_{ij}\ket{\psi_j},\qquad g=\frac{-i}{R_0}
\left(
  \begin{array}{cc}
    R_3+iR_4 & R_1+iR_2 \\
    R_1-iR_2 & -R_3+iR_4 \\
  \end{array}
\right)
\ee
Here $R_0=\sqrt{R_1^2+R_2^2+R_3^2+R_4^2}$. It is easy to see that $g$ is unitary and det$g=1$, thus belongs to $SU(2)$. Now the Berry phase and Berry curvature are transformed as $A^g=g^{-1}Ag+g^{-1}dg$ and $F^g=g^{-1}Fg$. The second Chern number can also be understood from $g$. Let $\ket{\psi}$ and $\ket{\psi^g}$ be defined on two 4D discs $D_1$ and $D_2$. The boundary between $D_1$ and $D_2$ are 3D sphere $S^3$. We know Tr$F^2$ are closed, thus can be locally written as Tr$F^2=d\omega_3$. Here $\omega_3=\mbox{Tr}(AF-A^3/3)$ is Chern-Simons 3-form. $\omega_3$ is not gauge invariant and is transformed as $\omega_3^g=\omega-\mbox{Tr}(g^{-1}dg)^3+d\alpha_2$. Then the second Chern number is also given by
\be
c_2&=&\frac{1}{8\pi^2}\Big(\int_{D_1}d\omega_3+\int_{D_2}d\omega^g_3\Big)
=\frac{1}{8\pi^2}\Big(\int_{\p D_1}d\omega_3-\int_{\p D_2}d\omega^g_3\Big)\nonumber\\
&=&\frac{1}{24\pi^2}\int_{S^3}\mbox{Tr}\Big[(g^{-1}dg)^3\Big]
=-\frac{1}{2\pi^2}\int_{S^3}\frac{1}{R_0^4}\sum_{i=1}^4(-1)^{i-1}R_idR_1\cdots
dR_{i-1}dR_{i+1}\cdots dR_{4}=-1\nonumber
\ee
This explicitly shows that the Chern number is related to $\pi_3(S^3)$.

To show the relation with the instanton solution of Non-abelian gauge theory, we first map the northern half of $S^4$ to $\mathbb{R}^4$ by a stereographic projection as follows
\be
r_i=\frac{R}{R_5}R_i,\qquad\mbox{Inverse:}\quad
R_i=\frac{r_i}{\sqrt{R^2+r^2}}\quad R_5=\frac{R}{\sqrt{R^2+r^2}}
\ee
for $i=1,\cdots,4$. It maps the north pole of $S^4$ to the origin or $\mathbb{R}^4$ and maps the equator to the infinite boundary of $\mathbb{R}^4$. After this mapping, the Berry phase can be written as
\be
A_{\mu}=-i\frac{r^2 g^{-1}\p_{\mu}g}{R^2+r^2+R\sqrt{R^2+r^2}},\qquad
g=\frac{1}{r}(r_4+ir_i\sigma_i)
\ee
which can be verified that do not satisfies the self-adjoint condition $F^a_{ij}=\pm\frac12\epsilon_{ijkl}F^a_{kl}$. Since it has the same Chern number as an instanton, one can always find a continuous coordinate transformation to connect these two solutions. Here we can slightly generalize the above stereographic projection as
\be
R_i=f(r)r_i,\qquad R_5=\sqrt{R^2-f^2(r)r^2},\qquad i=1,\cdots,4
\ee
Then the resulting Berry phase is
\be
A_{\mu}=-i\frac{f^2(r) g^{-1}\p_{\mu}g}{R^2+R\sqrt{R^2-f^2(r)r^2}},\qquad
g=\frac{1}{r}(r_4+ir_i\sigma_i)
\label{Amu}
\ee
Then it is easy to verify that by taking $f(r)=\frac{\sqrt{r^2+2R^2}}{r^2+R^2}$, Eq. (\ref{Amu}) will become the standard instanton solution $A_{\mu}=-i\frac{r^2}{R^2+r^2}g^{-1}\p_{\mu}g$. This is closely parallel with the fact that the $U(1)$ Berry phase of a time reversal breaking two band model is the same as a magnetic monopole solution.

\section{Three orbital copper-oxygen model with spin current order}

We can apply the same method to understand the three orbital copper-oxygen model given by Hamiltonian $H=H_{KE}+H'$ as discussed in section \ref{spin}. Then the Hamiltonian can be written in the matrix form as
\be
H=\left(\begin{array}{ccc}
0 & iR_1 & -iR_2+a\\
-iR_1 & 0 & is_zR_3\\
iR_2+a & -is_zR_3 & 0
\end{array}\right)
\label{Hm}
\ee
with $R_1=\sin\frac{k_x}{2}$, $R_2=-\sin\frac{k_y}{2}$ and $R_3=r\cos\frac{k_x}{2}\cos\frac{k_y}{2}$, note that $R_3$ is always positive for $-\pi<k_x,k_y<\pi$. $s_z$ is the $z$ component spin. Here we also introduce a small positive constant $a$ in order to lift the degeneracy of spin up and spin down bands. With the spin degree of freedom, the Hamiltonian is actually a 6 by 6 matrix as
\be
H=\left(\begin{array}{ccc}
0 & iR_1 & -iR_2+a\\
-iR_1 & 0 & 0\\
iR_2+a & 0 & 0
\end{array}\right)\otimes I
+\left(\begin{array}{ccc}
0 & 0 & 0\\
0 & 0 & iR_3\\
0 & -iR_3 & 0
\end{array}\right)\otimes s_z
\ee
The energy band can be solved from the following cubic equation
\be
E^3-(a^2+R^2)E+2aR_1(s_zR_3)=0
\label{En}
\ee
Here $R=\sqrt{R_1^2+R_2^2+R_3^2}$. At general momentum, the last term of above equation is different for spin up and spin down, so the degeneracy is lifted at these momentum. Now consider the 4 TRIM points $(0,0)$, $(0,\pi)$, $(\pi,0)$ and $(\pi,\pi)$. At $(0,0)$, $R_1=0$, and at other 3 points, $R_3=0$. The last term of Eq. (\ref{En}) is always zero at TRIM points, thus the spin up and spin down bands degenerate at these points as required.

As we already know from \cite{yanheAHE}, the top band of the Hamiltonian Eq.(\ref{Hm}) with $a=0$ and $s_z=\pm1$
has Chern number $\pm1$ respectively. Since adding a constant will not change the Chern number, for nonzero $a$, the spin up and down of the top bands of Eq. (\ref{Hm}) still have the Chern number $\pm1$ respectively. As in the two band case, we want to show that the top band has nontrivial Chern parity by counting the number of zero pairs of the Pfaffian $P(\vk)$. Again, in this case, the Pfaffian is just a number $P(\vk)=\bra{u_{\ds}(\vk)}\Theta\ket{u_{\us}(\vk)}$.

Suppose the 3 bands are $E_{1,2,3}$. It is easy to verify for small $a$ the 3 bands are always separated by gaps. We will only consider the top band $E_3$. Since the Hamiltonian is traceless, $E_1+E_2+E_3=0$, thus we must have $E_3>0$.

As before, there are many ways to write the eigenstate, we can write it in a particular gauge as
\be
\ket{u_{s_z}(\vk)}\propto\left(\begin{array}{c}
aE_3-R_1(s_zR_3)-iE_3R_2 \\
i(E_3(s_zR_3)-aR_1)-R_1R_2 \\
E_3^2-R_1^2
\end{array}\right)
\label{usz}
\ee

The wavefunction $\ket{u_{s_z}}$ is well defined if its 3 components of $R_i$ are not equal to zero at the same time. As in the two band example, the zeros usually occur at singular points of the wave function. Therefore, we first consider the special momentum point $\vk^*$ where $\ket{u_{s_z=+1}}$ is not well defined or its 3 components are all zero. From the real part of 2nd component, we have $R_1R_2=0$. If we have $R_1=0$, then the 3rd component requires $E_3=0$ which is impossible, thus we must have $R_2=0$ or $k_y=0$. Suppose we consider the case that $R_1>0$ or $k_x>0$, then we have $E_3=R_1$. The imaginary part of 2nd component gives us $R_3=a$. Thus we find out that $\vk^*=(2\cos^{-1}a,\,0)$. Then Eq (\ref{En}) become
$$
E^3-(2a^2+R_1^2)E+2a^2R_1=0
$$
and one can verify that $E=R_1$ indeed satisfying this equation. In summary, at $\vk^*$, we have $R_2=0$, $R_3=a$ and $E_3=R_1$ and the Hamiltonian becomes
\be
H_{s_z=+1}=\left(\begin{array}{ccc}
0 & iR_1 & a\\
-iR_1 & 0 & ia\\
a & -ia & 0
\end{array}\right)
\ee
The 3 eigenvalues are $R_1$ and $\frac12(-R_1\pm\sqrt{R_1^2+8a^2})$. Clearly, if $R_1>0$ and $a$ is small positive number, $R_1$ is top band. $\frac12(-R_1+\sqrt{R_1^2+8a^2})\approx 0$ is the middle band. $\frac12(-R_1-\sqrt{R_1^2+8a^2})\approx-R_1$ is the lower band. So it is consistent with our previous assumption that $R_1$ is top band. The eigenstate corresponding to $E_3=R_1$ is
\be
\ket{u_{+}}=(i,\,1,\,0)^T
\ee

Now we can also compute the top eigenvalue of $H_{s_z=-1}$ at $\vk^*$. At this point, we find
\be
H_{s_z=-1}=\left(\begin{array}{ccc}
0 & iR_1 & a\\
-iR_1 & 0 & -ia\\
a & ia & 0
\end{array}\right)
\ee
The 3 eigenvalues from bottom to top are $-R_1$, $\frac12(R_1-\sqrt{R_1^2+8a^2})$and $\frac12(R_1+\sqrt{R_1^2+8a^2})$ respectively. So the top band is $E_3=\frac12(R_1+\sqrt{R_1^2+8a^2})$ and the corresponding eigenstate is
\be
\ket{u_{-}}=\left(\begin{array}{c}
2a^2+R_1^2+R_1\sqrt{R_1^2+8a^2}\\
-i(2a^2+R_1^2+R_1\sqrt{R_1^2+8a^2})\\
a(3R_1+\sqrt{R_1^2+8a^2})
\end{array}\right)
\ee
Then we find $\bra{u_{-}}\Theta\ket{u_{+}}\propto\bra{u_{-}}\Big(\ket{u_{+}}\Big)^*=0$. Thus $\pm\vk^*$ is pair of zero points of $P(\vk)$.

Now we show that this is the only pair of zeros of $P(\vk)$. Above we only discussed the special momentum point $\vk^*$ on which the wave function Eq (\ref{usz}) is not well defined. Other than these points, we can always use Eq (\ref{usz}). Denote the top band of $H_{s_z=1}$ to be $E_p$ and top band of $H_{s_z=-1}$ to be $E_n$. It is easy to find that
\be
&&P(\vk)=\bra{u_{-}(\vk)}\Theta\ket{u_{+}(\vk)}
\propto\bra{u_{-}(\vk)}(\ket{u_{+}(\vk)})^*=A+iB\\
&&A=a^2(E_pE_n-R_1^2)-R_1^2(E_p^2+E_n^2)+E_p^2E_n^2-E_pE_n(R_2^2+R_3^2)+R_1^2R^2\nonumber\\
&&B=2a(E_pE_n-R_1^2)R_2^2\nonumber
\ee
Now we want to show $A$ and $B$ cannot be zero at the same time.

First we consider the case that $R_2\neq0$. If we assume $B=0$, then this means $E_pE_n=R_1^2$.
We know that $E_{p,n}$ satisfy equations $E_{p,n}^3-(a^2+R^2)E_{p,n}\pm 2aR_1R_3=0$. Summing up these two equations and making use of the fact that $E_{p,n}$ are positive, we find the following identity
\be
E_p^2+E_n^2=E_pE_n+a^2+R^2
\label{iden1}
\ee
Combine with $E_pE_n=R_1^2$, we find $E_p^2+E_n^2=a^2+R^2+R_1^2$. Now plug all the above result into the expression of $A$, we find that $A=-R_1^2(a^2+R_2^2+R_3^2)$. We know that both $a^2+R_2^2+R_3^2>0$ and $R_1^2=E_pE_n>0$, thus $A\neq0$ in this case.

Second, we consider the case $R_2=0$. From identity Eq. (\ref{iden1}), we find that
\be
A=(E_pE_n)^2+(a^2-R_1^2-R_3^2)(E_pE_n)-2a^2R_1^2
\ee
We also have $E_{p,n}$ satisfying $E_{p,n}^2-(a^2+R^2)\pm 2aR_1R_3/E_{p,n}=0$. Subtracting these two equations, we find that $(E_p-E_n)=-2aR_1R_3\frac1{E_pE_n}$ Squaring the above equation and making use of Eq. (\ref{iden1}), we find another identity
\be
a^2+R^2-E_pE_n=(2aR_1R_3)^2\frac1{(E_pE_n)^2}
\label{iden2}
\ee
Now if we assume that $A=0$, then we find the following two equations
\be
(E_pE_n)^2+(a^2-R_1^2-R_3^2)(E_pE_n)-2a^2R_1^2=0 \label{eq1}\\
-(E_pE_n)^3+(a^2+R^2)(E_pE_n)^2-(2aR_1R_3)^2=0   \label{eq2}
\ee
Multiplying Eq. (\ref{eq1}) with $(E_pE_n)$ and adding it to Eq. (\ref{eq2}), we find
$(E_pE_n)^2=2R_1^2R_3^2+R_1^2(E_pE_n)$. Combining this result with Eq. (\ref{eq1}), we find
$(a^2-R_3^2)(E_pE_n-2R_1^2)=0$. Since we do not consider the special point which satisfy $R_2=0$ and $R_3=\pm a$, we find $E_pE_n=2R_1^2$. By plugging this result back to Eq. (\ref{eq1}), we will find a contradiction. Therefore in this case, we also have $A\neq0$.

In Summary, other than the special points, there is no more zeros of $P(\vk)$. Thus the $Z_2$ invariance is $-1$ indicating the QSH state. Similar to the two band model, the singular points of the eigenfanuction and their gauge dependence can be explicitly demonstrated by connecting the $Z_2$ invariant to the second order Chern number. But the calculation will be much more complicated than the two band case.

\section{Conclusion}

We have discussed the the possible QSH state in a three-band model motivated by the lattice of copper-oxygen planes of the high-temperature superconductors. The calculation of the Chern parity or $Z_2$ invariant is through a detailed analysis of the singular points of the eigenfunctions. We show that the the zeros of the Pfaffians of time reversal operator are coincident with the singular points of the eigenfunction. It is known through the dimensional reduction that the nontrivial Chern parity corresponds to the odd second order Chern number of a corresponding higher dimensional model. From a wave function point of view, we show that the geometric meaning of these topological invariant manifests in the obstruction which prevent us to define the wave function and Berry phase globally, just as the first order Chern number in the QAH state. An interesting fact is that the non-abelian Berry phase of this 4D hamiltonian can be mapped to a non-abelian gauge theory instanton.


\begin{thebibliography}{21}
\expandafter\ifx\csname natexlab\endcsname\relax\def\natexlab#1{#1}\fi
\expandafter\ifx\csname bibnamefont\endcsname\relax
  \def\bibnamefont#1{#1}\fi
\expandafter\ifx\csname bibfnamefont\endcsname\relax
  \def\bibfnamefont#1{#1}\fi
\expandafter\ifx\csname citenamefont\endcsname\relax
  \def\citenamefont#1{#1}\fi
\expandafter\ifx\csname url\endcsname\relax
  \def\url#1{\texttt{#1}}\fi
\expandafter\ifx\csname urlprefix\endcsname\relax\def\urlprefix{URL }\fi
\providecommand{\bibinfo}[2]{#2}
\providecommand{\eprint}[2][]{\url{#2}}

\bibitem[{\citenamefont{Thouless et~al.}(1982)\citenamefont{Thouless, Kohmoto,
  Nightingale, and den Nijs}}]{tknn}
\bibinfo{author}{\bibfnamefont{D.~J.} \bibnamefont{Thouless}},
  \bibinfo{author}{\bibfnamefont{M.}~\bibnamefont{Kohmoto}},
  \bibinfo{author}{\bibfnamefont{M.~P.} \bibnamefont{Nightingale}},
  \bibnamefont{and} \bibinfo{author}{\bibfnamefont{M.}~\bibnamefont{den Nijs}},
  \bibinfo{journal}{Phys. Rev. Lett.} \textbf{\bibinfo{volume}{49}},
  \bibinfo{pages}{405} (\bibinfo{year}{1982}).

\bibitem[{\citenamefont{Haldane}(1988)}]{haldane}
\bibinfo{author}{\bibfnamefont{F.~D.~M.} \bibnamefont{Haldane}},
  \bibinfo{journal}{Phys. Rev. Lett.} \textbf{\bibinfo{volume}{61}},
  \bibinfo{pages}{2015} (\bibinfo{year}{1988}).

\bibitem[{\citenamefont{Haldane}(2004)}]{haldaneahe}
\bibinfo{author}{\bibfnamefont{F.~D.~M.} \bibnamefont{Haldane}},
  \bibinfo{journal}{Phys. Rev. Lett.} \textbf{\bibinfo{volume}{93}},
  \bibinfo{pages}{206602} (\bibinfo{year}{2004}).

\bibitem[{\citenamefont{Nagaosa et~al.}(2010)\citenamefont{Nagaosa, Sinova,
  Onoda, MacDonald, and Ong}}]{nagaosaahereview}
\bibinfo{author}{\bibfnamefont{N.}~\bibnamefont{Nagaosa}},
  \bibinfo{author}{\bibfnamefont{J.}~\bibnamefont{Sinova}},
  \bibinfo{author}{\bibfnamefont{S.}~\bibnamefont{Onoda}},
  \bibinfo{author}{\bibfnamefont{A.~H.} \bibnamefont{MacDonald}},
  \bibnamefont{and} \bibinfo{author}{\bibfnamefont{N.~P.} \bibnamefont{Ong}},
  \bibinfo{journal}{Rev. Mod. Phys.} \textbf{\bibinfo{volume}{82}},
  \bibinfo{pages}{1539} (\bibinfo{year}{2010}).

\bibitem[{\citenamefont{Avron et~al.}(1983)\citenamefont{Avron, Seiler, and
  Simon}}]{avronseilersimon}
\bibinfo{author}{\bibfnamefont{J.~E.} \bibnamefont{Avron}},
  \bibinfo{author}{\bibfnamefont{R.}~\bibnamefont{Seiler}}, \bibnamefont{and}
  \bibinfo{author}{\bibfnamefont{B.}~\bibnamefont{Simon}},
  \bibinfo{journal}{Phys. Rev. Lett.} \textbf{\bibinfo{volume}{51}},
  \bibinfo{pages}{51} (\bibinfo{year}{1983}).

\bibitem[{\citenamefont{He et~al.}(2012)\citenamefont{He, Moore, and
  Varma}}]{yanheAHE}
\bibinfo{author}{\bibfnamefont{Y.}~\bibnamefont{He}},
  \bibinfo{author}{\bibfnamefont{J.}~\bibnamefont{Moore}}, \bibnamefont{and}
  \bibinfo{author}{\bibfnamefont{C.~M.} \bibnamefont{Varma}},
  \bibinfo{journal}{Phys. Rev. B} \textbf{\bibinfo{volume}{85}},
  \bibinfo{pages}{155106} (\bibinfo{year}{2012}).

\bibitem[{\citenamefont{Varma}(1997)}]{cmv97}
\bibinfo{author}{\bibfnamefont{C.~M.} \bibnamefont{Varma}},
  \bibinfo{journal}{Phys. Rev. B} \textbf{\bibinfo{volume}{55}},
  \bibinfo{pages}{14554} (\bibinfo{year}{1997}).

\bibitem[{\citenamefont{Simon and Varma}(2002)}]{simon-cmv}
\bibinfo{author}{\bibfnamefont{M.~E.} \bibnamefont{Simon}} \bibnamefont{and}
  \bibinfo{author}{\bibfnamefont{C.~M.} \bibnamefont{Varma}},
  \bibinfo{journal}{Phys. Rev. Lett.} \textbf{\bibinfo{volume}{89}},
  \bibinfo{pages}{247003} (\bibinfo{year}{2002}).

\bibitem[{\citenamefont{Varma}(2006)}]{cmv06}
\bibinfo{author}{\bibfnamefont{C.~M.} \bibnamefont{Varma}},
  \bibinfo{journal}{Phys. Rev. B} \textbf{\bibinfo{volume}{73}},
  \bibinfo{pages}{155113} (\bibinfo{year}{2006}).

\bibitem[{\citenamefont{Bourges and Sidis}(2011)}]{bourges}
\bibinfo{author}{\bibfnamefont{P.}~\bibnamefont{Bourges}} \bibnamefont{and}
  \bibinfo{author}{\bibfnamefont{Y.}~\bibnamefont{Sidis}},
  \bibinfo{journal}{Comptes Rendus Physique} \textbf{\bibinfo{volume}{12}},
  \bibinfo{pages}{461} (\bibinfo{year}{2011}).

\bibitem[{\citenamefont{Kaminski et~al.}(2002)\citenamefont{Kaminski,
  Rosenkranz, Fretwell, Campuzano, Li, Raffy, Cullen, You, Olson, Varma
  et~al.}}]{kaminski}
\bibinfo{author}{\bibfnamefont{A.}~\bibnamefont{Kaminski}},
  \bibinfo{author}{\bibfnamefont{S.}~\bibnamefont{Rosenkranz}},
  \bibinfo{author}{\bibfnamefont{H.~M.} \bibnamefont{Fretwell}},
  \bibinfo{author}{\bibfnamefont{J.~C.} \bibnamefont{Campuzano}},
  \bibinfo{author}{\bibfnamefont{Z.}~\bibnamefont{Li}},
  \bibinfo{author}{\bibfnamefont{H.}~\bibnamefont{Raffy}},
  \bibinfo{author}{\bibfnamefont{W.~G.} \bibnamefont{Cullen}},
  \bibinfo{author}{\bibfnamefont{H.}~\bibnamefont{You}},
  \bibinfo{author}{\bibfnamefont{C.~G.} \bibnamefont{Olson}},
  \bibinfo{author}{\bibfnamefont{C.~M.} \bibnamefont{Varma}},
  \bibnamefont{et~al.}, \bibinfo{journal}{Nature (London)}
  \textbf{\bibinfo{volume}{416}}, \bibinfo{pages}{610} (\bibinfo{year}{2002}).

\bibitem[{\citenamefont{Li et~al.}(2008)\citenamefont{Li, Baledent, Barisic,
  Bourges, Y.~Cho, Sidis, Yu, Zhao, and Greven}}]{li}
\bibinfo{author}{\bibfnamefont{Y.}~\bibnamefont{Li}},
  \bibinfo{author}{\bibfnamefont{V.}~\bibnamefont{Baledent}},
  \bibinfo{author}{\bibfnamefont{N.}~\bibnamefont{Barisic}},
  \bibinfo{author}{\bibfnamefont{P.}~\bibnamefont{Bourges}},
  \bibinfo{author}{\bibfnamefont{B.~F.} \bibnamefont{Y.~Cho}},
  \bibinfo{author}{\bibfnamefont{Y.}~\bibnamefont{Sidis}},
  \bibinfo{author}{\bibfnamefont{G.}~\bibnamefont{Yu}},
  \bibinfo{author}{\bibfnamefont{X.}~\bibnamefont{Zhao}}, \bibnamefont{and}
  \bibinfo{author}{\bibfnamefont{M.}~\bibnamefont{Greven}},
  \bibinfo{journal}{Nature (London)} \textbf{\bibinfo{volume}{455}},
  \bibinfo{pages}{372} (\bibinfo{year}{2008}).

\bibitem[{\citenamefont{Avron et~al.}(1989)\citenamefont{Avron, Sadun, Segert,
  and Simon}}]{Simon}
\bibinfo{author}{\bibfnamefont{J.~E.} \bibnamefont{Avron}},
  \bibinfo{author}{\bibfnamefont{L.}~\bibnamefont{Sadun}},
  \bibinfo{author}{\bibfnamefont{J.}~\bibnamefont{Segert}}, \bibnamefont{and}
  \bibinfo{author}{\bibfnamefont{B.}~\bibnamefont{Simon}},
  \bibinfo{journal}{Commun. Math. Phys.} \textbf{\bibinfo{volume}{124}},
  \bibinfo{pages}{595} (\bibinfo{year}{1989}).

\bibitem[{\citenamefont{Kane and Mele}(2005)}]{kane-mele}
\bibinfo{author}{\bibfnamefont{C.~L.} \bibnamefont{Kane}} \bibnamefont{and}
  \bibinfo{author}{\bibfnamefont{E.~J.} \bibnamefont{Mele}},
  \bibinfo{journal}{Phys. Rev. Lett.} \textbf{\bibinfo{volume}{95}},
  \bibinfo{pages}{145802} (\bibinfo{year}{2005}).

\bibitem[{\citenamefont{Hasan and Kane}(2010)}]{hasankane}
\bibinfo{author}{\bibfnamefont{M.~Z.} \bibnamefont{Hasan}} \bibnamefont{and}
  \bibinfo{author}{\bibfnamefont{C.~L.} \bibnamefont{Kane}},
  \bibinfo{journal}{Rev. Mod. Phys.} \textbf{\bibinfo{volume}{82}},
  \bibinfo{pages}{3045} (\bibinfo{year}{2010}).

\bibitem[{\citenamefont{Moore}(2010)}]{moorenature}
\bibinfo{author}{\bibfnamefont{J.~E.} \bibnamefont{Moore}},
  \bibinfo{journal}{Nature (London)} \textbf{\bibinfo{volume}{464}},
  \bibinfo{pages}{194} (\bibinfo{year}{2010}).

\bibitem[{\citenamefont{Qi and Zhang}(2011)}]{qizhangreview}
\bibinfo{author}{\bibfnamefont{X.-L.} \bibnamefont{Qi}} \bibnamefont{and}
  \bibinfo{author}{\bibfnamefont{S.-C.} \bibnamefont{Zhang}},
  \bibinfo{journal}{Rev. Mod. Phys.} \textbf{\bibinfo{volume}{83}},
  \bibinfo{pages}{1057} (\bibinfo{year}{2011}).

\bibitem[{\citenamefont{Qi et~al.}(2008)\citenamefont{Qi, Hughes, and
  Zhang}}]{zhang}
\bibinfo{author}{\bibfnamefont{X.~L.} \bibnamefont{Qi}},
  \bibinfo{author}{\bibfnamefont{T.~L.} \bibnamefont{Hughes}},
  \bibnamefont{and} \bibinfo{author}{\bibfnamefont{S.~C.} \bibnamefont{Zhang}},
  \bibinfo{journal}{Phys. Rev. B} \textbf{\bibinfo{volume}{78}},
  \bibinfo{pages}{195424} (\bibinfo{year}{2008}).

\bibitem[{\citenamefont{Fukui et~al.}(2008)\citenamefont{Fukui, Fujiwara, and
  Hatsugai}}]{hatsugai}
\bibinfo{author}{\bibfnamefont{T.}~\bibnamefont{Fukui}},
  \bibinfo{author}{\bibfnamefont{T.}~\bibnamefont{Fujiwara}}, \bibnamefont{and}
  \bibinfo{author}{\bibfnamefont{Y.}~\bibnamefont{Hatsugai}},
  \bibinfo{journal}{J. Phys. Soc. Jpn.} \textbf{\bibinfo{volume}{77}},
  \bibinfo{pages}{123705} (\bibinfo{year}{2008}).

\bibitem[{\citenamefont{Witten}(1982)}]{Witten}
\bibinfo{author}{\bibfnamefont{E.}~\bibnamefont{Witten}},
  \bibinfo{journal}{Phys. Lett. B} \textbf{\bibinfo{volume}{117}},
  \bibinfo{pages}{324} (\bibinfo{year}{1982}).

\bibitem[{\citenamefont{Nakahara}(2003)}]{nakahara}
\bibinfo{author}{\bibfnamefont{M.}~\bibnamefont{Nakahara}},
  \emph{\bibinfo{title}{Geometry, Topology and Physics}}
  (\bibinfo{publisher}{Institute of Physics Publishing},
  \bibinfo{address}{Bristol}, \bibinfo{year}{2003}), \bibinfo{edition}{2nd} ed.

\end{thebibliography}
\end{document}